# Phenomenological Scaling Crossovers in Non-Equilibrium Critical Dynamics

Rong Li[1,2*], Qirui Ding [1,2], Weicheng Cui[1,2*]

[1] Research Center for Industries of the Future, Westlake University, Hangzhou, Zhejiang 310030, China

[2] Key Laboratory of Coastal Environment and Resources of Zhejiang Province, School of Engineering, Westlake University, Hangzhou, Zhejiang 310030, China.

[*] lirong13@westlake.edu.cn, cuiweicheng@westlake.edu.cn

## Abstract

Dynamical universality plays a fundamental role in understanding the scaling properties of critical dynamics, including absorbing phase transitions and physical aging. Although individual universality classes have been extensively studied, the theoretical framework for scaling crossovers between distinct dynamical regimes remains underdeveloped. In this work, we propose a phenomenological approach to describe dynamic scaling crossovers in reaction-diffusion systems, utilizing a Bernoulli differential equation with time-dependent reaction coefficients. Under a new scaling hypothesis for spatial fluctuations, we analytically derive five universal power-law crossover functions. These solutions accurately describe experimental observations of absorbing phase transitions in turbulent liquid crystals and reaction-diffusion crossovers in exciton-exciton recombination, revealing possible mechanisms associated with many-body correlations and two-component dynamics. These findings suggest the existence of universal scaling laws governing the transition between different dynamical states, motivating further studies using field-theoretic approaches and renormalization group analyses.

## 1 Introduction

Scaling laws and universality are central to understanding phase transitions and critical dynamics, as they elucidate how microscopically distinct systems, featuring different components and interactions, exhibit identical critical behaviors [1-3]. In far-from-equilibrium contexts, several prominent dynamical universality classes have been identified [4-7], including Kardar–Parisi–Zhang (KPZ) dynamical scaling of driven interfaces [8], diffusion-limited kinetics of reaction-diffusion systems [3, 9], and the Direct Percolation (DP) class of absorbing phase transitions [5, 6]. Over the past decade, these classes have been unambiguously verified in experiments across various physical contexts, ranging from turbulent liquid crystals to exciton recombination in low-dimensional materials [10-18].

However, many experimental studies reveal a more complex dynamical evolution that extends beyond a single universality class. Specifically, scaling crossovers—transitions between different scaling regimes—have been observed in multiple systems. For instance, exciton-exciton recombination in a laser-pumped carbon nanotube displays a crossover behavior from a reaction-limited decay following $1/t$ to a subsequent diffusion-limited scaling $t^{-1/2}$ [19]. Similarly, absorbing phase transition experiments in turbulent liquid crystals exhibit a multi-stage relaxation of active density and survival probability, evolving from an initial slow (slower than the linear time dependence of mean-field type) decay to a DP-class power-law relaxation, and ultimately reaching either an exponential decay or a plateau-like state [5, 6]. A significant theoretical gap remains in quantifying and understanding these scaling crossovers. Specifically, the conventional scaling ansatz $t^a F(\epsilon t^{1/\nu})$ for the DP class relaxation fails at early times, leading to divergences [5, 6]. Additionally, the observed reaction-diffusion crossover is significantly more abrupt than predicted by existing theories and numerical simulations of reaction-diffusion systems [19]. These discrepancies suggest that existing models may be insufficient away from the asymptotic limit [19]. Thus, an intriguing question arises regarding the underlying dynamic mechanism that drives these multi-state crossovers, and whether universality can emerge in the dynamic transition between different scaling classes remains an open challenge.

Generally, scaling crossover across multiple dynamical states in distinct time regimes may stem from transformations of the underlying dynamical mechanisms. External forces and internal dynamics drive these nonequilibrium systems far away from thermodynamic equilibrium states [4, 5, 20]. In the context of ultrafast photo-induced phase transitions [21-23], intense ultrafast laser pulses drive quantum matter far from equilibrium, not only increasing the temperature but also inducing significant non-thermal changes. These non-thermal effects include the redistribution of elementary excitation populations, the dynamic modification of coupling strengths, and the resonant driving of the crystal lattice [19]. Such internal variations in dynamic mechanisms induce significant spatial fluctuations and temporal evolutions, extending beyond the scope of mean field models with constant coefficients and may necessitate the introduction of time-dependent coefficients. For instance, an exponential decaying "temperature deviation" was usually introduced to the time-dependent Ginzburg-Landau equation to describing the time dependence induced by ultrafast laser [24-27]. Additionally, the kinetics of diffusion-influenced reactions are often described by a time-dependent rate coefficient $k(t)$, which asymptotically approaches $a + b/t$ or $t^{-1/2}$ at long-time [28-30]. The latter represent a time-dependent exciton–exciton annihilation rate, characterizing the one-dimensional diffusion mechanism of excitons. Nevertheless, a systematic theoretical framework describing the scaling crossover in dynamic state transitions remains to be established.

Here, as a first step, we develop a phenomenological approach for the dynamic scaling crossover that systematically addresses transitions between distinct scaling regimes in nonequilibrium phase transitions. This approach applies a moment model——a Bernoulli differential equation with time-dependent coefficients derived from higher-order spatial fluctuations in reaction-diffusion systems. By incorporating a simple scaling hypothesis for the time dependence, we derive the analytical solutions of these equations and identify five universal scaling functions that describe crossovers between stationary states and power-law or exponential relaxations. These results accurately capture experimental observations of absorbing transitions and exciton relaxations, unveiling the transition of the underlying dynamics mechanism and many-body correlations. Our findings suggest an extension to the standard framework of dynamic scaling from the well-established simple forms $t^\alpha$ to more general crossover functions, i.e., $[1 + (t/\tau)^m]^n f(|\epsilon|t^\beta)$, where $\tau$ is the characteristic crossover time, $\epsilon$ is the relative distance from the critical point, $m$, $n$ and $\beta$ are critical exponents, and $f$ is the universal scaling function.

# 2 A analytic theory of dynamic scaling crossover

## 2.1 Scaling crossover hypothesis

Near the critical point, order parameter fluctuations become strong and long-range, as indicated by the asymptotic divergence of the characteristic correlation length. Correspondingly, the characteristic relaxation time associated with the order parameter kinetics should also increase significantly. This critical slowing-down is described by an appropriate dynamic scaling hypothesis [4, 31]. For an off-critical infinitely large system, a two-parameter scaling law for the order parameter relaxation can be obtained as follows [5, 6, 32]:

$$\phi(\epsilon, t) = t^\alpha f(|\epsilon| t^\beta). \tag{1}$$

For critical relaxation ($\alpha < 0$) at the critical point of $\epsilon = 0$, the simple power law $\phi(0, t) = t^\alpha f(0)$ diverges as $t \to 0$, which is unphysical and inconsistent with experimental observations [17, 18]. This discrepancy suggests the simple power law may be insufficient away from the asymptotic limit of $t \to +\infty$ [17-19], necessitating an extension of the hypothesis to describe the overall dynamics. Thus, following the invariance principle under scaling transformations [5], we propose a revised hypothesis of dynamic scaling crossover as

$$\phi(t) = g(t^\alpha/\tau^\alpha) f(|\epsilon| t^\beta). \tag{2}$$

Here, the simple power function $t^\alpha$ in the dynamic scaling hypothesis is transformed into a scaling crossover function $g(t^\alpha/\tau^\alpha)$. This to-be-determined function represents a crossover from a constant at $t = 0$ to a function of $t^\alpha$ as $t \to +\infty$. In the following sections, we focus exclusively on the scaling crossover of this single-time scaling, deferring the investigation of the two-time correlations associated with physical aging to future work.

Generally, the diversities and complexities of non-equilibrium phase transitions suggest various forms of the function $g$. However, the present work focuses mainly on the general nonlinear dynamics at the critical point, given by $d\phi/dt = q(t)\phi^N$, and its extensions. Here, $q(t)$ represents a time-dependent extension of the reaction constant $q_0$, and $N \neq 1$ is a constant. The simplest scale-invariant expression of the reaction coefficient is $q(t) = ct^k$, where $c < 0$. In this

context, a straightforward scaling crossover function for $k \neq -1$ can be derived as follows:

$$\phi(t) = [(1-N)\int q(t)dt]^n = \phi(0)\left[1 + \left(\frac{t}{\tau}\right)^m\right]^n, \qquad (3)$$

where $m = 1 + k$ and $n = 1/(1-N)$ are two exponents, $\tau = \{\phi_0^{1/n} mn/c]\}^{1/m}$ is the characteristic crossover time. Eq. (3) represents a crossover from a constant $g(0)$ to a power-law scaling of $t^{mn}$ across $\tau$. This crossover may be universal for the nonlinear critical dynamics at the critical point. For instance, the critical solution $\phi \propto (1 + t/\tau)^{-1}$ of the mean-field model for binary coalescence and biological growth [4, 5] belong to this crossover class with $m = 1$, which corresponds to constant reaction rate. The diffusion-limited reaction and fractal reaction kinetics often correspond to $m = 1/2$ [14] and other $m \neq 1$ classes [13]. Interestingly, as observed in previous studies of turbulent boundary layers [33] and the resistivity of high-temperature superconductors [34], a functional form like Eq. (3) provides a universal description of scaling crossover, even though the variables involved are distance and temperature, respectively. In the following, we will derive the analytic function for both the power-law crossover form of Eq. (3) and the complete solution $\phi(t) = g(t)f(t)$ for the phase transition of the reaction-diffusion system, and verify their universalities with experimental observations.

## 2.2 Derivation of Bernoulli differential equation for a reaction-diffusion system

Let us consider a reaction-diffusion equation as follows [4]:

$$\frac{d\psi(\mathbf{x},t)}{dt} = -R[\psi(\mathbf{x},t)] + D\nabla^2\psi(\mathbf{x},t) + \xi(\mathbf{x},t), \qquad (4)$$

where $\psi(\mathbf{x},t)$ represents the density of reactive particles, $R[\psi(\mathbf{x},t)]$ denotes the reaction functional with a finite limit for $R[\psi]/\psi$, $D$ is the diffusion constant, and $\xi(\mathbf{x},t)$ represents all other fast degrees of freedom and internal reaction noise. In the vicinity of the extinction threshold, we may approximate the reaction functionals using a Taylor expansion: $R[\psi] = a\psi + \sum_i b_i \psi^i$, where $i \geq 2$ is a positive integer. We regard the mean density averaged over the entire system as the order parameter, $\phi(t) = \langle\psi\rangle$. Assuming that there is no net flux on the boundary and the multiplicative stochastic noise has a vanishing mean, a dynamic equation for the order parameter can be readily obtained in the form of an ordinary differential equation:

$$\frac{d\phi(t)}{dt} = -a\phi(t) - \sum_i b_i \langle \psi^i \rangle, \tag{5}$$

where $\langle \psi^{i+1} \rangle = \int \psi^{i+1}(x,t) d^d x / \int d^d x$, $d$ is the spatial dimensionality.

When diffusive mixing is strong enough to wipe out spatially correlated structures produced by the reactions, a mean-field approximation is expected to hold [5]. Thus, Eq. (5) can be simplified to a mean-field equation $d\phi/dt = -a\phi - \sum_i b_i \phi^i$. However, numerical simulations of non-equilibrium phase transitions reveal strong correlation effects in the critical region, and therefore the mean-field model without fluctuations is inadequate and cannot yield precise quantitative results [6]. Defining the spatial fluctuations as $\Delta = \psi - \phi$, we obtain $\langle \psi^i \rangle = \sum_{j=0}^{i} C_i^j \langle \Delta^{i-j} \rangle \phi^j = \phi^i + i \langle \Delta^{i-1} \rangle \phi + \sum_{j=2}^{i-2} C_i^j \langle \Delta^{i-j} \rangle \phi^j + \langle \Delta^i \rangle$ by utilizing $\langle \Delta \rangle = 0$. Therefore, we obtain an expansion for the order parameter dynamics as follows:

$$\frac{d\phi}{dt} = -\left[a + \sum_i i b_i \langle \Delta^{i-1} \rangle \right] \phi - \sum_i b_i \left[ 1 + \langle \left(\frac{\Delta}{\phi}\right)^i \rangle + \sum_{j=2}^{i-2} C_i^j \langle \left(\frac{\Delta}{\phi}\right)^{i-j} \rangle \right] \phi^i. \tag{6}$$

Generally, these spatial fluctuations are inherently time-dependent, revealing that the reaction coefficient of either linear or nonlinear terms in Eq. (6) is time-dependent. Consequently, a corresponding Bernoulli differential equation [35] for the nonlinear relaxation dynamics can be defined as:

$$\frac{d\phi}{dt} = p(t)\phi + q(t)\phi^N. \tag{7}$$

where $N$ can be any real number not equal to one, and the two time-dependent reaction coefficients are defined as:

$$p(t) = -a - z(t), \tag{8}$$

$$q(t) = -\sum_i b_i \left[ 1 + \sum_{j=0}^{i-2} C_i^j \langle \left(\frac{\Delta}{\phi}\right)^{i-j} \rangle \right] \phi^{i-N} + z(t)\phi^{1-N}, \tag{9}$$

It is straightforward to verify that Eq. (7) is equivalent to Eq. (6) for any real variable $z(t)$. By substituting $d\phi/dt$ with $d\phi^{1-N}/dt$ and applying integration by parts [36], we obtain the exact analytical solution of Eq. (7):

$$\phi(t) = e^{\int p(t)dt} \left\{ c_0 + (1-N) \int_{t_0}^{t} e^{(N-1)\int p(t)dt} q(t)dt \right\}^{\frac{1}{1-N}}, \tag{10}$$

where $c_0$ is a constant dependent on the initial time $t_0$.

Equation (10) provides the analytical solution for the dynamics of $\phi(t)$, whose specific form is determined by the explicit time dependence of $p(t)$ and $q(t)$. According to Eqs. (5), (8), and (9), in the absence of variations in external fields, the reaction rates $a$ and $b$ remain constant. In this context, the only source of nontrivial time dependence of $p(t)$ and $q(t)$ arises from the terms associated with non-zero spatial fluctuations $\langle \Delta^{i-j} \rangle$. These fluctuations are determined by the specific physical setup: bulk properties, boundary, and initial conditions. In contrast, when varying external fields are applied—e.g., laser irradiation in photo-induced dynamics [26, 27], or electric and magnetic quenches in liquid crystal and Bose-Einstein condensate transitions [37]—these reaction rates $a$ and $b$ also become time-dependent.

## 2.3 Universality classes of power-law scaling crossover

Let us discuss the solution of Eq. (10) for specific physical scenarios. The simplest case is the mean-filed approximation of the reaction-diffusion model [5], given by $\dot{\phi} = p_0 \phi + q_0 \phi^N$, which represents a combination of one-body and $N$-body reactions. Away from the critical point of $p_0 = 0$, the analytical solution of this equation is $\phi(t) = [A + (\phi_0^{1-N} - A)e^{p_0 t/n}]^n$, where $A = -q_0/p_0$, $\phi_0$ is the initial value at $t = 0$, and $n = 1/(1-N)$. At the critical point, where $p_0 = 0$, the solution becomes $\phi(t) = \left\{ \phi_0^{1/n} + q_0 t/n \right\}^n$. It is evident that there is a crossover from the initial constant $\phi_0$ to the power-law scaling $t^n$ in the large $t$ limit. However, these mean-field solutions are expected to be valid only when diffusive mixing is strong enough to eliminate spatially correlated structures, typically dimensions above 4 [5]. Therefore, it is necessary to consider the time dependence of $p(t)$ and $q(t)$ for critical dynamics in lower-dimensional systems (typically $d \leq 4$).

For nonequilibrium phase transitions, the time dependence of $p(t)$ and $q(t)$ arises from both internal physical configurations and external field influences, exhibiting diverse forms. However, the universality hypothesis posits that critical dynamics should fall into specific universality classes. Consequently, a crucial scaling hypothesis for rate coefficients is proposed: for typical classes of critical reaction dynamics, scale invariance constraints both spatial fluctuations and mean-filed behavior; thus, both the linear and nonlinear reaction rates should follow a simple power-law time evolution, as follows,

$$p(t) \to p_0 + bt^h \quad \text{and} \quad q(t) \to q_0 + ct^k, \qquad (11)$$

at the short- or long-time limit, where $p_0$, $q_0$, $b$, $c$, $h$, and $k$ are constants.

Firstly, we consider the relaxation dynamics at the critical point ($p = 0$) with a power-law dependence for the many-body reaction, given by $q(t) = \sum_i b_i \langle \psi^i \rangle \phi^{-N} = ct^k$. In this context, the analytical solution for Eq. (10) is

$$\phi = \phi_0 \left[ 1 + \left(\frac{t}{\tau}\right)^m \right]^n, \qquad (12)$$

where $m = 1 + k$, $n = 1/(1 - N)$, and $\tau = \left\{\phi_0^{1/n} mn/c\right\}^{1/m}$ is the characteristic crossover time. In fact, this power-law time dependence of the rate coefficient applies in many nonclassical situations [13], including fractal reaction kinetics, coalescence in one dimension, and A+B reactions on a square lattice. Therefore, Eq. (12) is a notable universal expression that extends the mean-field scaling (corresponding to $m = 1$), namely, the scaling-crossover function for the critical relaxation, as demonstrated by experiment observations in Sec. 3.

Secondly, we consider the relaxation dynamics away from the critical point ($p(t) \neq 0$) while maintaining the power-law dependence for the many-body reaction $q(t) = ct^k$. To specify $p(t)$, we recall that the kinetics for a class of diffusion-influenced reactions are described by a time-dependent rate coefficient, with its long-term behavior approaching $p(t) = a + b/t$, reflecting the diminishing likelihood of reactant encounters over time due to diffusion constraints and the progressive consumption of reactants [28, 29]. For the simplest case of the irreversible coagulation process, $a = 0$ [38]. In this context, the long-term behavior can be determined by power-law time-dependent linear and nonlinear reaction coefficients of $p(t) = b/t$ and $q(t) = ct^k$, where $k \neq b(1 - N) - 1$. Thus, the solution of this system is given by:

$$\phi(t) = At^b \left\{1 + \left(\frac{t}{\tau}\right)^m\right\}^n, \tag{13}$$

where $A$ is a constant related to the initial value, $m = b(N-1) + k + 1$, and $\tau = \{A^{1/n}mn/c\}^{1/m}$. Eq. (13) represents a dynamic scaling crossover from $t^b$ to $t^{b+mn}$ across the characteristic crossover time $\tau$, which will be demonstrated by experimental observations of exciton-exciton recombination shown in Section 3.

Thirdly, we address the nonphysical divergence of $p(t) = b/t \to \infty$ as $t \to 0$, since any realistic reaction should not exhibit an infinite reaction rate. There are two realistic $t \to 0$ limits, i.e., $p(t) \to 0$ and finite value, which can be expressed as $p(t) \to p_0 + bt^h$ with $h \geq 0$. One of the simplest power-law scaling function that satisfies both the zero- and long-time limits is $p(t) = f(t)b/t$, where $f(t) = t^m/(\tau^m + t^m)$, with $m \geq 1$ and $\tau > 0$. As time increases from zero to positive infinity, $f(t)$ transitions from $(t/\tau)^m$ to $1$, and thus the rate coefficient $p(t)$ transitions from $bt^{m-1}/\tau^m$ to $b/t$. Additionally, if the short-time evolution of the nonlinear coefficient also has a power-law multiplier $f(t)$, the nonlinear dynamics equation should be

$$\frac{d\phi}{dt} = \frac{b}{t}f(t)\phi + ct^k f^L(t)\phi^N, \tag{14}$$

where $L$ is a real number. For the special case of $m = b(N-1)/L$, the solution of this equation is

$$\phi(t) = \phi_0 \left[1 + \left(\frac{t}{\tau}\right)^m\right]^{\frac{b}{m}} \left\{1 + \left(\frac{t}{\tau^*}\right)^{m^*}\right\}^n, \tag{15}$$

where $\tau^* = \tau\left[\phi_0^{1/n}m^*n/c\tau^{k+1}\right]^{1/m^*}$ and $m^* = b(N-1) + k + 1$. Equation (15) represents a dynamic scaling crossover between $\phi_0$, $\propto t^b$ and $\propto t^{b+m^*n}$.

Fourthly, consider a pure linear reaction with a finite constant rate at the $t \to 0$ limit, expressed as $p(t) = p_0 + f(t)b/t$. The solution to this dynamic is given by

$$\phi = \phi_0 \left[1 + \left(\frac{t}{\tau}\right)^m\right]^{b/m} e^{p_0 t}. \tag{16}$$

Interestingly, this corresponds to the Bernoulli differential equation with a finite linear reaction $p(t) = p_0 \neq 0$ and an exponential nonlinear reaction coefficient $q(t) = ct^k \exp[(1-N)p_0 t]$, which is likely to be present in fluctuation $\langle \Delta^j/\phi^N \rangle$ terms if $\Delta$ exhibits weaker exponential decay than the mean value.

**Table I. Functional forms of dynamic scaling crossover associated with power-law.** $p(t)$ and $q(t)$ are time-dependent linear and nonlinear rate coefficients in the Bernoulli differential equations $\dot{\phi} = p(t)\phi + q(t)\phi^N$, respectively. Here, $n = 1/(1-N)$, $N$ is the nonlinear index, $m = 1 + k$ for the first and third cases, $K$ is a positive integer, and $f(t) = t^m/(\tau^m + t^m)$ is a time-dependent factor. For simplicity, $A$ and $B_i$ are used to replace the complex constant factors in the final solutions.

| $p(t)$ | $q(t)$ | $\phi(t)$ |
|---|---|---|
| $0$ | $ct^k$ | $\phi_0\left[1+\left(\frac{t}{\tau}\right)^m\right]^n$ |
| $p_0$ | $ct^K$ | $\left\{Ae^{p_0 t} + \sum_{i=0}^{K} B_i\, t^{K-i}\right\}^n$ |
| $\frac{b}{t}$ | $ct^k$ | $At^b\left[1+\left(\frac{t}{\tau}\right)^m\right]^n$ |
| $p_0 + \frac{b}{t}f(t)$ | $0$ | $\phi_0 e^{p_0 t}\left[1+\left(\frac{t}{\tau}\right)^m\right]^{\frac{b}{m}}$ |
| $\frac{b}{t}f(t)$ | $ct^k f^L(t)$ | $\phi_0\left[1+\left(\frac{t}{\tau}\right)^m\right]^{\frac{b}{m}}\left[1+\left(\frac{t}{\tau^*}\right)^{m^*}\right]^n$ |

Fifthly, another interesting crossover behavior occurs with a constant one-body rate coefficient and time-dependent nonlinear interactions, i.e., $p(t) = p_0$ is a constant, and $q(t) = ct^k$. This captures dynamics in systems with constant exponential growth or decay and power-law time evolution of many-body reactions. Utilizing the general analytical solution Eq. (10), one can obtain

$$\phi(t) = e^{p_0 t}\{A + bn^k p_0^{-m}\Gamma(m, p_0 t/n)\}^n, \qquad (17)$$

where $A$ is a constant determined by the initial value, $n = 1/(1-N)$, and $m = 1 + k$, where $\Gamma(s, x) = \int_0^x \tau^{s-1} e^{-\tau} d\tau$ is the lower incomplete gamma function. When $k = K$ is a positive integer, a simple explicit solution is obtained as

$$\phi(t) = \left\{Ae^{p_0 t/n} + \sum_{i=0}^{K} B_i\, t^{K-i}\right\}^n, \qquad (18)$$

where $B_i = -bn^i K!/(K-i)!\, p_0^{i+1}$. Eq. (18) represents a crossover between $\phi(t) \propto e^{p_0 t}$ and

$\phi(t) \propto t^{Kn}$. Table I summarizes the five different functional forms of scaling crossover obtained in our phenomenological approach.

## 3 Scaling crossover of relaxation kinetics in reacting particle systems

Recently, prominent power-law scaling has been experimentally observed in two nonstationary relaxation kinetics of reacting particle systems. The first class involves the active-to-absorbing state phase transition with DP scaling, observed in intermittent ferrofluidic spikes [16], electro-hydrodynamic convection in thin nematic liquid crystals [17, 18], and non-equilibrium phase transition on quantum computers [39]. Another class is characterized by anomalous diffusion-limited power-law decay in one dimension, evident in the exciton recombination kinetics in molecular wires [13], certain polymer chains [14], and carbon nanotubes [15, 40]. This section will demonstrate that these two classes serve as excellent platforms for verifying our scaling crossover theory.

### 3.1 Absorbing phase transitions in turbulent liquid crystals

Absorbing states, which systems may enter but never escape, and phase transitions into these states are expected to be ubiquitous across phenomena such as forest fires, epidemics, and spatiotemporal chaos. Theoretical and simulation studies have established that most absorbing phase transitions exhibit the same critical behavior, forming the DP universality class [6]. Experimentally, the DP critical exponents have been unambiguously verified at the transition between different turbulent states in nematic liquid crystals [17, 18]. However, previous studies [18, 39] have shown that the DP scaling functions deviate from the experimental observations of the initial relaxation of turbulent liquid crystal and the ultimate transition in quantum computers. For example, the theoretical prediction of the functional $t^{-\alpha}F(\epsilon t^{1/\nu})$ derived from the conventional scaling ansatz and numerical simulation of the contact process diverge early on [5, 6]. This discrepancy arises from the oversight of the slow dynamic crossover from the fully invaded state to a power-law critical decay in turbulent liquid crystals.

Here, we demonstrate that absorbing phase transitions possess universal scaling functionals and have a simple analytical form. In Sec. 2.3, we derived the analytical solution $\phi = \phi_0[1+(t/\tau)^m]^n$ (Eq. (12)) for the nonlinear relaxation dynamic $d\phi/dt = ct^k \phi^N$ at the critical point. As shown in Fig. 1 a and b, this scaling-crossover function is confirmed by experimental relaxation data of turbulent liquid crystals at the critical voltage. Equation (12) precisely characterizes the overall crossover dynamics of both the active density and survival probability, from a slow decay near the invaded state to a power-law critical decay $\propto t^{-\alpha}$. In contrast, as shown in Fig. A2, the prediction with $m \sim 1$ close to the mean-field solution $\phi(t) = \{1 + t/\tau\}^n$ for $d\phi/dt = q_0 \phi^N$ deviates from the initial experimental observation, representing an overly abrupt crossover of the nonlinear reaction $\phi^N$ compared to $ct^k \phi^N$. This may demonstrates that the power-law time dependence of the nonlinear reaction rate, which originates from spatial fluctuations, is crucial for slow critical relaxations.

The scaling crossover formula $[1+(t/\tau)^m]^n$ enables us to update the conventional simple scaling ansatzes $t^{-\alpha} F(\epsilon t^{1/\nu})$ [6, 17, 18] to a new scaling crossover functional for the overall dynamics near the critical point, as follows,

$$\phi(t) = \phi_0 \left[1+\left(\frac{t}{\tau}\right)^m\right]^n F\left(\epsilon t^{\frac{1}{\nu}}\right). \tag{19}$$

where $\alpha$ and $\nu$ are critical exponents, $\epsilon$ is the deviation of the controlling parameter from criticality, and $F(\epsilon t^{1/\nu})$ is the universal scaling function. For the nematic liquid crystal shown in Fig. 1 [17], $\epsilon = (V^2/V_c^2 - 1)$, where $V$ and $V_c$ are the applied and critical voltage on the liquid crystal, respectively. Eq. (19) implies that the time series for different voltages collapse onto a single curve $F\left(\epsilon t^{\frac{1}{\nu}}\right)$ when $(\phi/\phi_0)[1+(t/\tau)^m]^n$ is plotted as a function of $|\epsilon|^\nu t$. As shown in Fig. 2 (c) and (d), the data for survival probability and order parameter do collapse reasonably well, with the upper and lower branches corresponding to $V > V_c$ and $V < V_c$, respectively. In this collapse, the critical exponents $m = \nu = 1.3$ and $mn = 0.433$ agree with the DP exponents [17, 18]. It is noteworthy that the characteristic crossover times $\tau = 0.25$ and 2.5 seconds are constants in both cases. These findings suggest that the universality of absorbing phase transitions across multiple time scales not only exists but also can be quantified by the phenomenological scaling crossover function described in Eq. (19). However, further theoretical work, including renormalization group analysis and numerical validation, is necessary to demonstrate the universality of these crossovers.

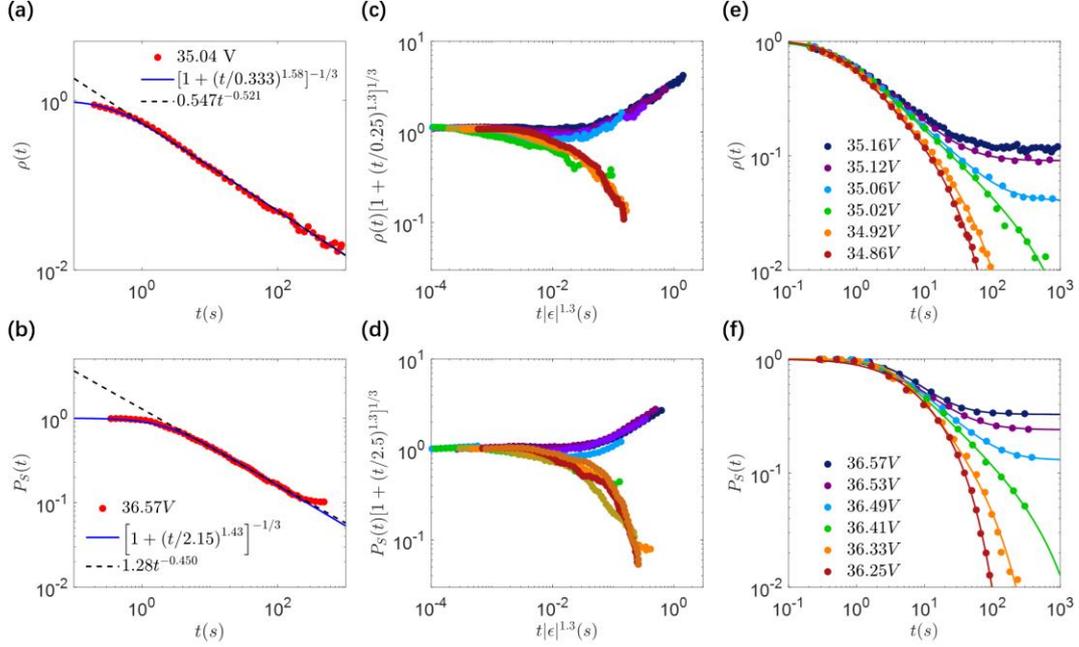

**FIG. 1 Dynamic transition of turbulent states in nematic liquid crystal after the critical quench.**
(a) Symbols represent the order parameter (the ratio of active sites to total sites) at the critical voltage $V_c = 35.04$ V [18]. (b) Symbols represent the survival probability (the probability that a cluster starting from a single active seed survives) at the critical voltage $V_c = 36.45$ V [18]. The solid blue lines indicate the fitting using the scaling crossover formula Eq. (12) with $n = -1/3$ and adjustable parameters $m$ and $\tau$. The dashed black line represents the fitting of conventional scaling $\propto t^{-\alpha}$ with $\alpha$ close to the DP scaling exponent $0.50 \pm 0.08$ [17]. (c) and (d) show experimental data [18] with rescaled axes $|\epsilon|^\nu t$ and $\phi(t)[1+(t/\tau)^m]^{-n}$, where $\nu = 1.3$ and $\tau = 0.25$s (c) and 2.5s (d), demonstrating data collapsing. (e) and (f) display the same data for the order parameter and survival probability below and above the critical voltages. The solid blue lines indicate the specific scaling-crossover function of Eq. (15) and (16) with $\phi_0 = 1$, $n = -1/3$ and adjustable parameters $m$, $\tau$, and $\tau^*$, see the Appendix A.

The Bernoulli differential equation also enables us to derive the explicit analytic expressions for the overall dynamics of $\phi(t)$ both above and below the critical region. As observed, the active density decays exponentially with a characteristic time for lower voltages and converges to a finite value for higher voltages [17, 18]. Therefore, in the simplest case, the dynamics below the critical point exhibit similar asymptotic behaviors described by Eq. (16) $\phi = \phi_0[1 + (t/\tau)^m]^n e^{-t/\tau^*}$ at both the $t \to 0$ and $t \to +\infty$ limits, where $\tau^* = \tau_- \epsilon^{-\nu}$. Above the critical voltage, the relaxation dynamics also exhibit similar asymptotic behaviors described by Eq. (15) $\phi = \phi_0[1 + (t/\tau)^m]^n [1 + (t/\tau^*)^m]^{-n}$ at both the $t \to 0$ and $t \to +\infty$ limits, where $\tau^* = \tau_+ \epsilon^{-\nu}$. Indeed, as shown in Fig. 2 (e) and (f), Eq. (15) and (16) agree well with the experimental data for active density and survival probability.

Interestingly, the quantification of observed data from universal functionals, as described by Eqs. (12), (15), and (16), reveal the many-body effects underlying the initial slow decay dynamics. Upon comparison in Fig. A1, it is evident that the four-body model with $n = 1(1 - N) = 1/3$ describes experimental data well. In contrast, the two-body models exhibit obvious deviations in the early stages. This highlights the necessity of incorporating many-body (rather than two-body) effect to explain the dynamics. This many-body effect is also quantitatively valid in the data collapse and overall dynamics of $\phi(t)$ both above and below the critical region, as shown in Fig. 2 (c)-(f). This suggests that the percolation dynamics near the totally invaded state in turbulent liquid crystals are likely dominated by many-body effects rather than conventional two-body reactions.

These findings reveal that the dynamical universality of absorbing phase transitions, driven by a class of time-dependent spatial fluctuations and many-body correlations, is governed by the scaling crossover functional Eq. (12): (a) scaling crossover induced by spatial fluctuations in the critical regime exhibits an invariant two-stage critical dynamics, evolving from an initial slow relaxation to the universal DP scaling of $t^{-\alpha}$; (b) the universal scaling functional above and below the critical point, Eq. (19), exhibits a three-stage relaxation process governed by a power-law dependence on the control parameter deviation. This discovery also paves the way for further theoretical advancements in absorbing phase transitions: the analytical solutions enable precise characterization of absorbing phase transitions, shedding new insights into the underlying mechanisms and offering new perspectives (e.g., incorporating many-body reaction) for studying field-theoretic descriptions of DP near criticality [41].

## 3.2 Two-component relaxation in exciton-exciton recombination

In exciton-exciton recombination, a pair of singlet excitons exchange energy: one is eliminated while the other transitions to a high-energy state, subsequently losing energy to the lattice and returning to its original state: $A + A \rightarrow A^* \rightarrow A + \text{heat} \rightarrow A$. According to existing theories [40], this coalescence is approximated as a sequential diffusion and reaction process, with the coalescence time $-ndt/dn$, predicted to be the sum of the time to the first encounter and the additional time required for multiple reaction attempts, expressed as $t_s = a/n^2 + b/n$, where $n$ represents the exciton density, and $a$ and $b$ are constants. Qualitatively, this theory predicts a crossover from reaction-limited behavior ($n \sim t^{-1}$) at early times to diffusion-limited kinetics ($n \sim t^{-1/2}$) at later times. However, quantitatively, the experimentally observed crossover in exciton-exciton recombination within carbon nanotubes [40] is significantly more abrupt than that predicted by the sequential diffusion-reaction model, as illustrated in Fig. 2 (a) and (b). This suggests anomalous reaction kinetics in a low-dimensional nonequilibrium stochastic system, indicating that there is some exotic mechanism need to clarified.

Interestingly, our theory of dynamic scaling crossover provides a general solution Eq. (13) for the power-law to power-law crossover. As shown in Fig. 2 a and b, both the density decay and the corresponding exponent of the exciton-exciton recombination data are accurately described by the power-law crossover formula

$$n = \frac{c_1}{t_s}\left[1 + \left(\frac{t_s}{c_2}\right)^4\right]^{\frac{1}{8}}, \qquad (20)$$

where the corresponding exponent is $\alpha' = (t_s/n)dn/dt_s$, with $c_1 = 0.3$ and $c_2 = 3.7$.

As we know, Eq. (13) provides the simplest and precise description for the experimentally observed reaction-diffusion crossover. Thus, it enables further exploration of the underlying dynamic mechanism. As detailed in Sec. 2.3, our theory predicts that a power-law to power-law crossover corresponds to the Bernoulli differential equation, expressed as $dn/dt = bn/t + ct^k n^N$.

Meanwhile, Sec. 2.2 demonstrates that the spatial fluctuations is one of the source for the time-dependent reaction in reaction-diffusion systems. Therefore, we conjecture that spatial fluctuations or inhomogeneities may be the critical source for the reaction-diffusion crossover in exciton-exciton recombination. In this context, the two-term differential equation suggests a simple two-fluid crossover model for understanding the reaction-diffusion crossover: assume that excitons within a single carbon nanotube are approximately segregated into high-density and low-density groups. The high-density excitons, constituting a fraction $f(t)$, undergo reaction-limited coalescence, while the low-density excitons, constituting a fraction $1 - f(t)$, undergo diffusion-limited coalescence. Notably, as the exciton density decreases, the proportion of high-density excitons may decrease from $f(t) \to 1$ as $t \to 0$ to $f(t) \to 0$ as $t \to +\infty$.

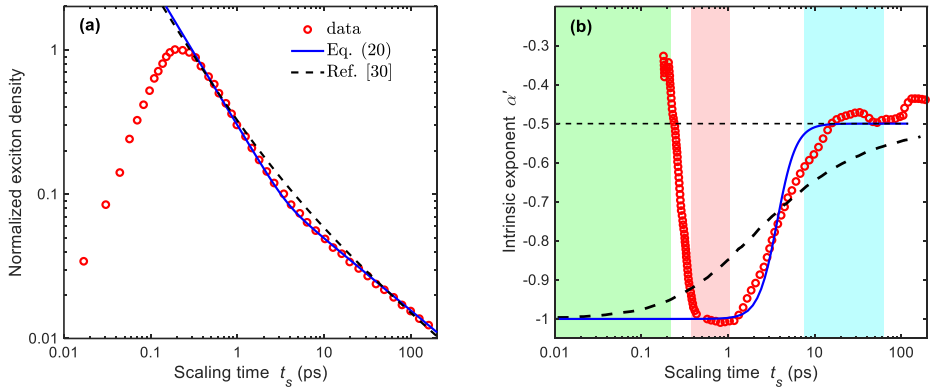

**FIG. 2 Reaction-diffusion crossover for exciton-exciton recombination in carbon nanotube.** (a) Normalized exciton density (symbols) for 104 nJ pump pulses [40], plotted against the scaling time $t_s$. The solid blue lines indicate our solution of the power-law crossover. Eq. (20), with $c_1 = 0.3$ and $c_2 = 3.7$. While the dashed black line represents the prediction of the sequential diffusion-reaction model $n = 2n_2/(\sqrt{1 + 4t_s/t_2} - 1)$ [40], where $t_2 = 3.8 \pm 0.2$ ps and $n_2 = 0.07$. (b) Decay exponent $\alpha' = (t_s/n)dn/dt_s$ for the data in Fig. 2a. The red symbols show the exponent determined from experimental data, and the solid blue line represents our scaling crossover prediction ($\alpha' = -1 + 1/2[1 + (3.7/t_s)^4]$) from Eq. (20), while the dashed black line is the prediction $\alpha' = (-1 + 2t_s/t_2 + \sqrt{1 + 4t_s/t_2})/(1 + 4t_s/t_2 + \sqrt{1 + 4t_s/t_2})$ of the sequential diffusion-reaction model [40].

We examine the coalescence dynamics of two coupled components within a single carbon nanotube. Initially, based on effective rate equations for reaction-limited and diffusion-limited processes, the independent effective rate equations for the high-density ($n_A$) and low-density ($n_B$) groups are given by: $dn_A/dt = \alpha f(t)^2 n^2$ and $dn_B/dt = \beta[1 - f(t)]^3 n^3$, where $n_A = f(t)n$, $n_B = [1 - f(t)]n$. Thus, the decay dynamics of the total excitation density are governed by $dn/dt = \alpha[f(t)]^2 n^2 + \beta[1 - f(t)]^3 n^3$. However, complete spatial separation of high and low-density regions is unrealistic, leading to coupling between the two components. This introduces additional terms, such as $\alpha^* f(t)[1 - f(t)]n^2$, $\beta_1[f(t)]^2[1 - f(t)]n^3$, and $\beta_2 f(t)[1 - f(t)]^2 n^3$. Consequently, the reaction-limited coalescence terms become $[\alpha^* + (\alpha - \alpha^*)f(t)]f(t)n^2$, while the diffusion-limited coalescence terms are $\{\beta + (\beta + \beta_1 - \beta_2)[f(t)]^2 + (\beta_2 - 2\beta)f(t)\}[1 - f(t)]n^3$. Given the partial spatial separation of exciton types, the probability of inter-species coalescence is lower than intra-species, i.e., $0 < -\alpha^* < -\alpha$ and $0 < -\beta_i < -\beta$, $i = 1$ and $2$. Hence, it can be demonstrated that: $n^3 \alpha f n^2 + \beta(1 - f)n^3 < dn/dt < \alpha f^2 n^2 + \beta(1 - f)^3$. The upper and lower bounds of the exciton decay rates are specific cases of the effective kinetic equation:

$$\frac{dn}{dt} = \alpha f^\gamma n^2 + \beta(1 - f)^\eta n^3, \tag{21}$$

where $(\gamma, \eta) = (1,1)$ and $(2,3)$ represent these limits, respectively. Therefore, we can use this framework, associated with $1 < \gamma < 2$ and $1 < \eta < 3$, to approximate the dynamics of exciton decay in a single carbon nanotube.

As shown in Appendix C, we demonstrated that any dynamic equation of the form $d\phi/dt = g(\phi)\phi^K + h(\phi)\phi^L$ must correspond to a Bernoulli differential equation Eq. (7) and has a general solution as given by Eq. (10). The power-law crossover solution, Eq. (20), can be transformed to the form $n = (c_1/\sqrt{c_2 t})\,[1 + (c_2/t)^4]^{1/8}$. This solution also corresponds to a Bernoulli differential equation as follows,

$$\frac{dn}{dt} = -\frac{1}{2}\frac{n}{t} - \frac{1}{2}\frac{c_1^8}{t^9 n^7}. \tag{22}$$

This equation should be mapped to Eq. (21) with coefficients $f^\gamma$ and $(1 - f)^\eta$ evolving from 1 to 0 and 0 to 1, respectively. This constraint can be satisfied by transforming Eq. (22) to $dn/dt = -c_1^8 n^2/(t^9 n^9) - n^3(1 - c_1^8/t^8 n^8)/(2tn^2)$. Substituting the expression for $n(t)$ from Eq. (20)

into this equation, we can obtain Eq. (21) with the high-density exciton fraction given by

$$f(t) = \frac{1}{1 + (t/c_2)^4}, \tag{23}$$

and two exponents $\gamma = 9/8$, $\eta = 5/4$, and the coefficients for reaction $\alpha = -1/c_1$ and diffusion $\beta = -c_2/2c_1^2$. This fraction can also be expressed as a function of density $f(n) = 1 - 2/[\sqrt{1 + 4(nc_2/c_1)^8} + 1]$, indicating the fraction of low-density excitons is $2/[\sqrt{1 + 4(nc_2/c_1)^8} + 1]$. It can be readily verified that this function $f(t)$ and the exponents $\gamma$ and $\eta$ satisfy the physical constraints: $f(t)$ decaying from 1 to 0 as $t$ increases, $1 < \gamma < 2$ and $1 < \eta < 3$, strongly supporting the validity of our coupled two-component model. Interestingly, the nonlinear exponents $(\gamma, \eta) = (9/8, 5/4)$ are close to the upper limits $(1,1)$, indicating significant coupling (or mixing) between these two components.

Our two-component model predicts the time evolution of exciton density ratios. Specifically, Eq. (23) predicts that, in the long-time limit for times much greater than $c_2 = 3.7$ ps, the high-density exciton fraction adheres to a scaling law of $f(t) \propto t^{-4}$. Conversely, the fraction of low-density excitons scales $1 - f(t) \propto t^4$ as $t \to 0$. To validate these predictions, we recommend employing spatially resolved experimental techniques to image exciton densities in real space, such as spatially resolved electron-energy-loss spectroscopy [42]. Based on detailed calculations using experimental parameters [40], the inter-exciton spacing for high-density excitons under reaction-limited coalescence is $\leq 1.5$ nm. Therefore, we suggest statistically analyzing the proportion of high-density excitons with predicted spacings $\leq 1.5$ nm to test whether their evolution follows the $f(t)$ function. This will further elucidate the presence of two exciton types in carbon nanotubes and confirm whether their decay behaviors obey the predicted laws. Additionally, it is intriguing to determine whether this power-law scaling of the fraction evolution and the exponent 4 universal are universal for the critical dynamics of exciton-exciton recombination or if they are highly dependent on the nature of substrate material (i.e., element and dimension) and the pumping laser (i.e., frequency and power). Therefore, systematic experiments on the temporal evolution of exciton densities for different substrate materials and pumping lasers are expected to yield valuable insights and warrant further experimental investigation.

Furthermore, utilizing Eqs. (23) and (21), we can deduce the evolution equation governing the fraction of high-density excitons as follows,

$$\frac{df}{dt} = -\frac{4}{c_1} f^\gamma (1-f) n. \tag{24}$$

Eqs. (21) and (24) constitute the complete set of coupled differential equations for the dynamics of the two-component system, clearly elucidating the physical mechanisms of system relaxation. For instance, the nearly equal nonlinear exponents $(\gamma, \eta) \sim (1,1)$ of Eq. (21) suggests a significant coupling between the two components, while Eq. (24) highlights that this coupling predominantly governs the decay of the high-density fraction. Similarly, whether Eqs. (21) and (24) also apply to other two-component critical dynamics is an interesting problem.

However, directly establishing and solving such complex coupled differential equations can be challenging due to the coefficients involving $f$ or $n$. In contrast, expressing the differential equations for $f$ and $n$ as independent Bernoulli equations with time-dependent coefficients, i.e., $dn/dt = -n/2t - c_1^8/(2t^9 n^7)$ and $df/dt = -4f(1-f)/t$, allows us to derive exact analytic solutions. These solutions enable precise characterization and a quantitative understanding of the experimental observations. This analysis underscores that our theory of Bernoulli equations with time-dependent coefficients and their universal scaling crossover solutions is not only crucial for inferring the coupled equations from experimental data but also a powerful tool for solving complex coupled dynamics in multi-component systems.

## 4  Conclusion

To conclude, we have uncovered a possible critical universality in the dynamic crossover from one scaling to another for nonequilibrium relaxations, such as active-to-absorbing phase transitions in turbulent liquid crystal and exciton-exciton recombination in carbon nanotubes. This universal scaling crossover (e.g., $[1 + (t/\tau)^m]^n f(|\epsilon| t^\beta)$) arises from the power-law time evolutions of the linear and nonlinear reaction rates, as given by Eq. (11), $p(t) \to bt^h$ and $q(t) \to ct^k$. As demonstrated in Sec. 2.2 and 2.3, these power laws step from the scale invariance of spatial fluctuations $-\sum_i \sum_{j=0}^{i} b_i C_i^j \langle \Delta^{i-j} \rangle \phi^{j-N} \propto t^k$. Therefore, we recommend future experiments and simulations to systematically measure the time scaling for higher-order moments of spatial fluctuations in physical aging, absorbing phase transition systems, and reaction-diffusion systems. These investigations will not only corroborate our pivotal scaling assumption but may also reveal the new universality of spatial fluctuations.

Interestingly, the temporal counterparts of these fluctuations, known as Reynolds stresses in turbulence, have been demonstrated to play a critical role in determining the scaling crossover of transport in the turbulent boundary layer [33, 43]. Both strands of research highlight the fundamental importance of understanding the universal dynamics of higher-order fluctuations in nonequilibrium systems. Particularly, if the evolution of these fluctuations deviates from simple power laws, it may open the possibility of uncovering a novel universal class distinct from those presented in this work. It is worth mentioning that the scaling-crossover functionals summarized in Table I are obtained through our phenomenological approach in the present work. Whether these correspond to distinct universality classes in the renormalization group sense remains to be established. Therefore, we suggest that field-theoretic approaches using stochastic Langevin equations and renormalization group flow should be applied to different physical contexts of reaction-diffusion systems to demonstrate the universality of these scaling crossovers.

A further extension to our theoretical model, such as the scaling forms of time-dependent coefficients in the Bernoulli differential equation, can yield intriguing findings regarding the dynamical universality of diverse biological systems. For example, in biological growth or population dynamics, environmental changes and intrinsic factors, such as genetic mutations and aging, significantly influence metabolic processes and growth rates. This results in phenomena such as exponential growth shifts in bacterial populations [44], multi-stage tumor development [45, 46], and the occurrence of seasonal or multiple pandemic waves [47, 48]. We find that the evolution of these systems may correspond to a new universal class characterized by exponential scaling crossover. Exponential and logistic time dependencies can describe this behavior in the linear and nonlinear reaction rates of the Bernoulli differential equation, which will be detailed in our subsequent publications. Therefore, our work provides a new possibility for studying the universality of critical scaling crossover in non-equilibrium phenomena.

# Acknowledgment

The authors acknowledge Z. S. She for inspiring R. Li to study the scaling transition. This research was supported by the Scientific Research Funding Project of Westlake University under Grant No. WU2024A001. The manuscript's language has been polished using the O3-mini-high models. The authors meticulously reviewed the manuscript for accuracy and completeness, assuming full responsibility for its integrity.

# Reference

[1] L.P. Kadanoff, W. Götze, D. Hamblen, R. Hecht, E. Lewis, V.V. Palciauskas, M. Rayl, J. Swift, D. Aspnes, J. Kane, Static phenomena near critical points: theory and experiment, Rev. Mod. Phys., 39 (1967) 395.
[2] P.C. Hohenberg, B.I. Halperin, Theory of dynamic critical phenomena, Rev. Mod. Phys., 49 (1977) 435-479.
[3] U.C. Täuber, Critical dynamics: a field theory approach to equilibrium and non-equilibrium scaling behavior, Cambridge University Press2014.
[4] U.C. Täuber, Phase Transitions and Scaling in Systems Far from Equilibrium, Annu. Rev. Condens. Matter Phys., 8 (2017) 185-210.
[5] H. Hinrichsen, Non-equilibrium critical phenomena and phase transitions into absorbing states, Adv. Phys., 49 (2000) 815-958.
[6] M.H.H.S. Lübeck, Non-equilibrium phase transitions, Vol. I: Absorbing Phase Transitions, Springer, Dordrecht, 2008.
[7] M. Henkel, M. Pleimling, Non-Equilibrium Phase Transitions: Volume 2: Ageing and Dynamical Scaling Far from Equilibrium, Springer Science & Business Media2011.
[8] M. Kardar, G. Parisi, Y.-C. Zhang, Dynamic Scaling of Growing Interfaces, Phys. Rev. Lett., 56 (1986) 889-892.
[9] S.A. Rice, Diffusion-Limited Reactions, Elsevier Science, Amsterdam, 1985.
[10] P.L. Schilardi, O. Azzaroni, R.C. Salvarezza, A.J. Arvia, Validity of the Kardar-Parisi-Zhang equation in the asymptotic limit of metal electrodeposition, Phys. Rev. B, 59 (1999) 4638-4641.
[11] M. Myllys, J. Maunuksela, M. Alava, T. Ala-Nissila, J. Merikoski, J. Timonen, Kinetic roughening in slow combustion of paper, Phys. Rev. E, 64 (2001) 036101.
[12] K.A. Takeuchi, M. Sano, Universal Fluctuations of Growing Interfaces: Evidence in Turbulent Liquid Crystals, Phys. Rev. Lett., 104 (2010) 230601.
[13] R. Kopelman, Fractal Reaction Kinetics, Science, 241 (1988) 1620-1626.
[14] R. Kroon, H. Fleurent, R. Sprik, Diffusion-limited exciton fusion reaction in one-dimensional tetramethylammonium manganese trichloride (TMMC), Phys. Rev. E, 47 (1993) 2462-2472.
[15] R.M. Russo, E.J. Mele, C.L. Kane, I.V. Rubtsov, M.J. Therien, D.E. Luzzi, One-dimensional diffusion-limited relaxation of photoexcitations in suspensions of single-walled carbon nanotubes, Phys. Rev. B, 74 (2006) 041405.

[16] P. Rupp, R. Richter, I. Rehberg, Critical exponents of directed percolation measured in spatiotemporal intermittency, Phys. Rev. E, 67 (2003) 036209.
[17] K.A. Takeuchi, M. Kuroda, H. Chaté, M. Sano, Directed percolation criticality in turbulent liquid crystals, Phys. Rev. Lett., 99 (2007) 234503.
[18] K.A. Takeuchi, M. Kuroda, H. Chaté, M. Sano, Experimental realization of directed percolation criticality in turbulent liquid crystals, Phys. Rev. E, 80 (2009) 051116.
[19] A. de la Torre, D.M. Kennes, M. Claassen, S. Gerber, J.W. McIver, M.A. Sentef, Colloquium: Nonthermal pathways to ultrafast control in quantum materials, Rev. Mod. Phys. , 93 (2021) 041002.
[20] A. Onuki, Phase transition dynamics, Cambridge University Press, Cambridge 2002.
[21] C. Bao, P. Tang, D. Sun, S. Zhou, Light-induced emergent phenomena in 2D materials and topological materials, Nat. Rev. Phys., 4 (2022) 33-48.
[22] T. Dong, S.J. Zhang, N.L. Wang, Recent Development of Ultrafast Optical Characterizations for Quantum Materials, Adv. Mater., 35 (2022) e2110068.
[23] S. Koshihara, T. Ishikawa, Y. Okimoto, K. Onda, R. Fukaya, M. Hada, Y. Hayashi, S. Ishihara, T. Luty, Challenges for developing photo-induced phase transition (PIPT) systems: From classical (incoherent) to quantum (coherent) control of PIPT dynamics, Phys. Rep., 942 (2022) 1-61.
[24] R. Yusupov, T. Mertelj, V.V. Kabanov, S. Brazovskii, P. Kusar, J.-H. Chu, I.R. Fisher, D. Mihailovic, Coherent dynamics of macroscopic electronic order through a symmetry breaking transition, Nat. Phys., 6 (2010) 681-684.
[25] P. Beaud, A. Caviezel, S.O. Mariager, L. Rettig, G. Ingold, C. Dornes, S.W. Huang, J.A. Johnson, M. Radovic, T. Huber, T. Kubacka, A. Ferrer, H.T. Lemke, M. Chollet, D. Zhu, J.M. Glownia, M. Sikorski, A. Robert, H. Wadati, M. Nakamura, M. Kawasaki, Y. Tokura, S.L. Johnson, U. Staub, A time-dependent order parameter for ultrafast photoinduced phase transitions, Nat. Mater., 13 (2014) 923-927.
[26] J. Maklar, Y.W. Windsor, C.W. Nicholson, M. Puppin, P. Walmsley, V. Esposito, M. Porer, J. Rittmann, D. Leuenberger, M. Kubli, M. Savoini, E. Abreu, S.L. Johnson, P. Beaud, G. Ingold, U. Staub, I.R. Fisher, R. Ernstorfer, M. Wolf, L. Rettig, Nonequilibrium charge-density-wave order beyond the thermal limit, Nat. Commun., 12 (2021) 2499.
[27] S. Duan, W. Xia, C. Huang, S. Wang, L. Gu, H. Liu, D. Xiang, D. Qian, Y. Guo, W. Zhang, Ultrafast Switching from the Charge Density Wave Phase to a Metastable Metallic State in 1T-TiSe$_2$, Phys. Rev. Lett., 130 (2023) 226501.
[28] M.J. Potter, B. Luty, H.-X. Zhou, J.A. McCammon, Time-Dependent Rate Coefficients from Brownian Dynamics Simulations, J. Phys. Chem., 100 (1996) 5149-5154.
[29] H.-X. Zhou, A.J.B.J. Szabo, Theory and simulation of the time-dependent rate coefficients of diffusion-influenced reactions, Biophys. J., 71 (1996) 2440-2457.
[30] G. Soavi, S. Dal Conte, C. Manzoni, D. Viola, A. Narita, Y. Hu, X. Feng, U. Hohenester, E. Molinari, D. Prezzi, K. Müllen, G. Cerullo, Exciton–exciton annihilation and biexciton stimulated emission in graphene nanoribbons, Nat. Commun., 7 (2016) 11010.
[31] B.I. Halperin, D.R. Nelson, Resistive transition in superconducting films, J. Low Temp. Phys., 36 (1979) 599-616.
[32] T. Vicsek, F. Family, Dynamic Scaling for Aggregation of Clusters, Phys. Rev. Lett., 52 (1984) 1669-1672.
[33] Z.-S. She, X. Chen, F. Hussain, Quantifying wall turbulence via a symmetry approach: a Lie group theory, J. Fluid Mech., 827 (2017) 322-356.
[34] R. Li, Z.-S. She, Symmetry-constrained quantum coupling in non-Fermi-liquid transport, Chin. Phys.

B, 32 (2023) 067104.

[35] J. Bernoulli, Explicationes, annotationes et additiones ad ea quæin actis superiorum annorum de curva elastica, isochrona paracentrica, & velaria, hinc inde memorata, & partim controversa leguntur; ubi de linea mediarum directionum, aliisque novis, Acta Eruditorum Dec (1695) 537–553.

[36] A.E. Parker, Who solved the Bernoulli differential equation and how did they do it?, College Math. J., 44 (2013) 89-97.

[37] C. Chin, R. Grimm, P. Julienne, E. Tiesinga, Feshbach resonances in ultracold gases, Rev. Mod. Phys., 82 (2010) 1225-1286.

[38] D. ben-Avraham, M.A. Burschka, C.R. Doering, Statics and dynamics of a diffusion-limited reaction: Anomalous kinetics, nonequilibrium self-ordering, and a dynamic transition, Journal of Statistical Physics, 60 (1990) 695-728.

[39] E. Chertkov, Z. Cheng, A.C. Potter, S. Gopalakrishnan, T.M. Gatterman, J.A. Gerber, K. Gilmore, D. Gresh, A. Hall, A. Hankin, M. Matheny, T. Mengle, D. Hayes, B. Neyenhuis, R. Stutz, M. Foss-Feig, Characterizing a non-equilibrium phase transition on a quantum computer, Nat. Phys., 19 (2023) 1799-1804.

[40] J. Allam, M.T. Sajjad, R. Sutton, K. Litvinenko, Z. Wang, S. Siddique, Q.H. Yang, W.H. Loh, T. Brown, Measurement of a Reaction-Diffusion Crossover in Exciton-Exciton Recombination inside Carbon Nanotubes Using Femtosecond Optical Absorption, Phys. Rev. Lett., 111 (2013) 197401.

[41] L.T. Adzhemyan, M. Hnatič, E.V. Ivanova, M.V. Kompaniets, T. Lučivjanský, L. Mižišin, Field-theoretic analysis of directed percolation: Three-loop approximation, Phys. Rev. E, 107 (2023) 064138.

[42] L.H.G. Tizei, Y.-C. Lin, M. Mukai, H. Sawada, A.-Y. Lu, L.-J. Li, K. Kimoto, K. Suenaga, Exciton Mapping at Subwavelength Scales in Two-Dimensional Materials, Phys. Rev. Lett., 114 (2015) 107601.

[43] X. Chen, F. Hussain, Z.-S. She, Quantifying wall turbulence via a symmetry approach. Part 2. Reynolds stresses, J. Fluid Mech., 850 (2018) 401-438.

[44] M. Basan, T. Honda, D. Christodoulou, M. Hörl, Y.-F. Chang, E. Leoncini, A. Mukherjee, H. Okano, B.R. Taylor, J.M. Silverman, C. Sanchez, J.R. Williamson, J. Paulsson, T. Hwa, U. Sauer, A universal trade-off between growth and lag in fluctuating environments, Nature, 584 (2020) 470-474.

[45] S.L. Spencer, M.J. Berryman, J.A. García, D. Abbott, An ordinary differential equation model for the multistep transformation to cancer, J. Theor. Biol., 231 (2004) 515-524.

[46] I.A. Rodriguez-Brenes, N.L. Komarova, D. Wodarz, Tumor growth dynamics: insights into evolutionary processes, Trends Ecol. Evol., 28 (2013) 597-604.

[47] G. Chowell, M.A. Miller, C. Viboud, Seasonal influenza in the United States, France, and Australia: transmission and prospects for control, Epidemiol. Infect., 136 (2008) 852-864.

[48] T. Hale, N. Angrist, A.J. Hale, B. Kira, S. Majumdar, A. Petherick, T. Phillips, D. Sridhar, R.N. Thompson, S. Webster, Government responses and COVID-19 deaths: Global evidence across multiple pandemic waves, PloS one, 16 (2021) e0253116.

# Appendix A: Fitting parameters for the absorbing phase transitions

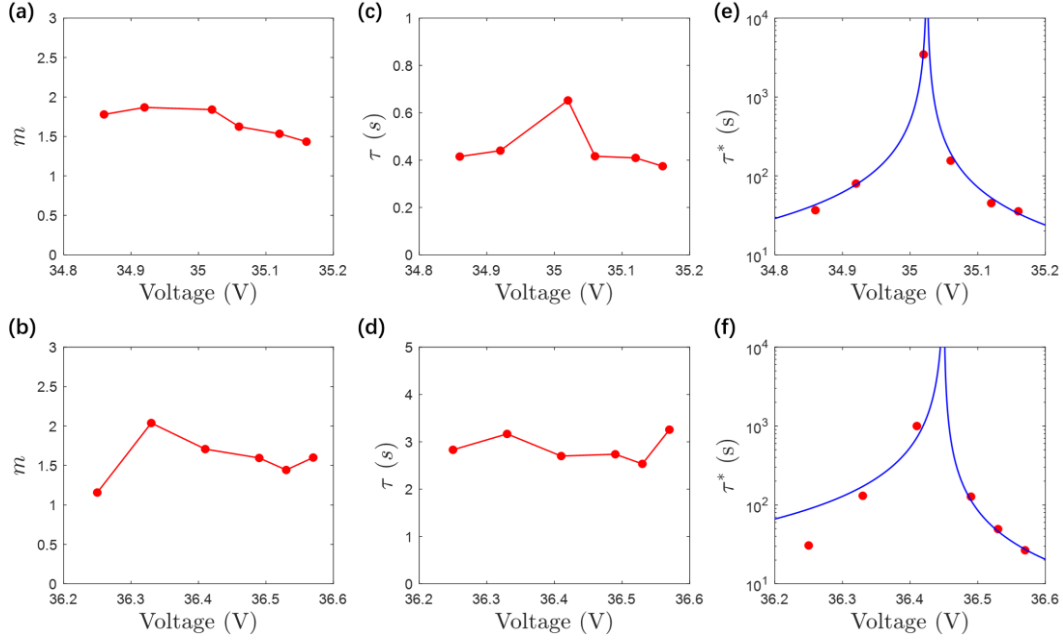

Fig. A1. Voltage dependence of the fitting parameters $m$, $\tau$, and $\tau^*$ for the solid lines shown in Fig.1 (e) and (f). (a), (c) and (e) show the parameter values for the order parameter (active density) in the voltage range 34.86–35.16 V. (b), (d) and (f) show the parameter values for the survival probability in the voltage range 36.25–36.57 V. The solid blue lines represent the fitting curves of $\tau^* = \tau_{\pm}(V^2/V_c^2 - 1)^{-\nu}$, with $V_c = 35.025$, $\tau_- = 0.1$s and $\tau_+ = 0.06$s for (e) and $V_c = 36.45$, $\tau_- = 0.25$ and $\tau_+ = 0.04$ for (f).

In Sec. 3.1, the asymptotic formulas $\phi = \phi_0[1 + (t/\tau)^m]^n e^{-t/\tau^*}$ and $\phi = \phi_0[1 + (t/\tau)^m]^n [1 + (t/\tau^*)^m]^{-n}$ are proposed phenomenally to describe the dynamics of absorbing phase transitions below and above the critical voltage, respectively. These formulas are used to fit the evolution data of the order parameter and survival probability in turbulent liquid crystals, as shown by the solid lines in Fig. 1 (e) and (f). In these fits, the data normalization requires $\phi_0 = 1$, and $n = -1/3$ is chosen based on the effective four-body reaction obtained from Fig. 1 (a)-(d). Here, in Fig. A1, we present the voltage dependence of the adjustable parameters $m$, $\tau$, and $\tau^*$ in the ranges 34.86–35.16 V for the order parameter and 36.25–36.57 V for the survival probability. As shown in (a) and (b), the exponent $m$ describing the crossover gradualness is nearly constant for both the active density (1.7 ± 0.2) and the survival probability (1.6 ± 0.3). While the characteristic crossover time $\tau = 0.45 \pm 0.1$ and $2.9 \pm 0.3$ are roughly constant in each case, they exhibit a magnitude difference for the active density and survival probability. As illustrated in (e) and (f), the second crossover time $\tau^*$ displays a divergent voltage dependence, following $(V^2/V_c^2 - 1)^{-\nu}$, as indicated by the solid lines, where $V_c$ represents the critical voltage, and $\nu = 1.3$ agrees with the DP exponents [17, 18].

## Appendix B: Model comparisons for many-body effects

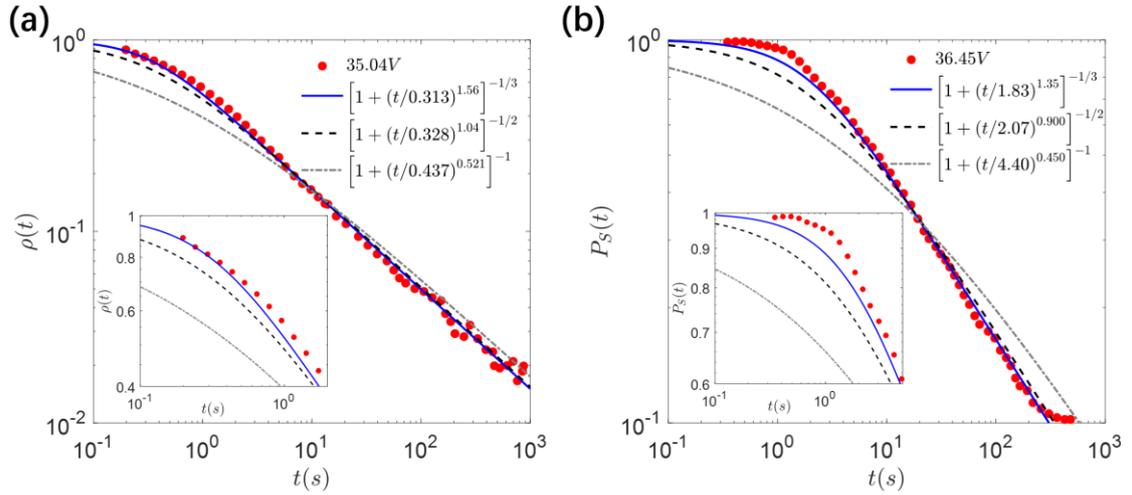

Fig. A2. Comparison of experimental data (symbols) with the scaling crossover function Eq. (12). Panels (a) and (b) depict the active density and survival probability data of the turbulent liquid crystal at the critical voltage [17, 18], respectively. The solid blue lines correspond to the four-body model ($n = -1/3$), whereas the dashed red and dotted green lines represent the three-body ($n = -1/2$) and two-body ($n = -1$) models, respectively. The values of $m = -0.521/n$ and $-0.45/n$ for the power-law regime was obtained in Fig. 1 (a) and (b). $\tau$ is derived from least-square fits to $\ln \phi$. The insets provide magnified views of the initial-stage dynamics.

The scaling crossover function $\phi(t) = \{1 + (t/\tau)^m\}^n$ of Eq. (12) applies to various many-body models with different $N$. However, as shown in Fig. A2, comparisons of the experimental data with different models reveal that the four-body reaction model ($N = 4$) provides the most accurate description of the active density and survival probability, capturing both the early slow relaxation and the later power-law decay. In contrast, as shown in the insets, the three-body ($N = 3$) and two-body ($N = 2$) models exhibit obvious deviations during the initial stages, underscoring the necessity of incorporating large $N$-body effect to explain the dynamics.

## Appendix C: Correspondence of dynamic equations

Bernoulli differential equations and their analytic solutions provide a novel approach to deriving the governing equations for nonequilibrium critical dynamics. Typically, nonequilibrium critical relaxation and absorbing phase transitions are self-organized phenomena, often described by time-independent self-organized dynamical equations. For instance, in the absence of spatial heterogeneity and noise, self-organized dynamics involving many-body processes can be transformed into a two-part dynamic described by the following order parameter dynamic equation:

$$\frac{d\phi}{dt} = g(\phi)\phi^K + h(\phi)\phi^L. \tag{A1}$$

Here, $g(\phi)$ and $h(\phi)$ are functions of order parameter $\phi$, determining the dynamical strength of each process. For Eq. (A1), it can be formally reduced to a Bernoulli-like differential equation, specifically, $\dot\phi = [g(\phi)\phi^{K-1}]\phi + h(\phi)\phi^L$. Since $\phi$ is a function of $t$, $g(\phi)\phi^{K-1}$ and $h(\phi)$ must also be functions of $t$. Therefore, Eq. (A1) can be transformed into this kind of Bernoulli differential equation with a formal solution as follows:

$$\phi(t) = e^{\int g(\phi)\phi^{K-1}dt}\left\{c + (1-N)\int_{t_0}^{t}e^{(N-1)\int g(\phi)\phi^{K-1}dt}h(\phi)dt\right\}^{\frac{1}{1-L}}. \tag{A2}$$

The mapping from Eq. (A1) to Eq. (7) provides a powerful tool for uncovering the dynamical mechanisms underlying the critical dynamics of complex systems. Consider a scenario where the many-body nature of two processes in a system, precisely the values of $K$ and $L$, is unknown, along with $g(\phi)$ and $h(\phi)$. If experiments exhibit typical scaling crossover behavior, we can derive $g(\phi)$ and $h(\phi)$ from the Bernoulli differential equation and its analytic solutions. For instance, if by fitting experimental data, we find that $\phi(t)$ conforms to one of the five universal scaling crossover types listed in Table I with acceptable accuracy, it may satisfy the Bernoulli differential equation in time-dimension, i.e., Eq. (7).

Comparing Eq. (7) and (A1), we find that $g(\phi)\phi^K + h(\phi)\phi^L = p(t)\phi + q(t)\phi^N$. Since the functions on either side of the equation may not necessarily correspond one-to-one, it is essential to map them based on their physical properties. Assuming a physical analysis allows for such a mapping with a function $y(t)\phi^M$ satisfying

$$g(\phi)\phi^K = p(t)\phi - y(t)\phi^M. \tag{A3}$$

Thus, we can find that

$$h(\phi)\phi^L = q(t)\phi^N + y(t)\phi^M. \tag{A4}$$

Therefore, the rate coefficients of the two potential functions are $g(\phi) = p(t)\phi^{1-K} - y(t)\phi^{M-K}$ and $h(\phi) = q(t)\phi^{N-L} + y(t)\phi^{M-L}$, respectively. Since $\phi = \mathcal{F}(t) \Rightarrow t = \mathcal{F}^{-1}(\phi)$, we have $p(t) = p[\mathcal{F}^{-1}(\phi)]$. Therefore, we can obtain $g(\phi)$ and $h(\phi)$ from the Bernoulli differential equation as

$$g(\phi) = p[\mathcal{F}^{-1}(\phi)]\phi^{1-K} - y[\mathcal{F}^{-1}(\phi)]\phi^{M-K}, \tag{A5}$$

$$h(\phi) = q[\mathcal{F}^{-1}(\phi)]\phi^{N-L} + y[\mathcal{F}^{-1}(\phi)]\phi^{M-L}. \tag{A6}$$

Therefore, we propose a three-step method to derive $g(\phi)$ and $h(\phi)$: Firstly, derive a dynamics equation for the order parameter that describes one or many body processes with rate coefficient $g(\phi)$ and $h(\phi)$ to be determined; Secondly, derive the scaling crossover solution $\phi(t)$ and the corresponding Bernoulli differential equation by best fitting the experimental data, as well as the function of $t = f^{-1}(\phi)$; Finally, define the function $y(t)\phi^M$, which satisfies the physical constraints of $g(\phi)$ and $h(\phi)$, and substitutes $t = f^{-1}(\phi)$ into Eq. (A3) and (A4) to obtain the formulas for $g(\phi)$ and $h(\phi)$ from Eq. (A5) and (A6). In all three steps, the derivations are not unique; thus, it is necessary to select the most realistic combination based on a comparison with reasonable physical mechanisms, simplicity, symmetry, and other guiding principles. Subsequently, we can study the underlying self-organizing mechanism from the $\phi$ or time dependence of $g(\phi)$ and $h(\phi)$. This method has been demonstrated to be useful in Sec. 3.2.